\documentclass[aps,twocolumn,letterpaper,superscriptaddress,showpacs,amsmath,amssymb]{revtex4-1}

\usepackage{graphicx}
\usepackage[utf8]{inputenc}
\usepackage{amsmath,amssymb,subfigure,color}
\usepackage{comment}

\begin{document}

\preprint{???}

\title{Detectability of Small-Scale Dark Matter Clumps with Pulsar Timing Arrays}

\author{Kazumi Kashiyama} 
\affiliation{Research Center for the Early Universe (RESCEU), Graduate School of Science, The University of Tokyo, Tokyo 113-0033, Japan}
\affiliation{Department of Physics, University of Tokyo, Tokyo 113-0033, Japan}

\author{Masamune Oguri} 
\affiliation{Research Center for the Early Universe (RESCEU), Graduate School of Science, The University of Tokyo, Tokyo 113-0033, Japan}
\affiliation{Department of Physics, University of Tokyo, Tokyo 113-0033, Japan}
\affiliation{Kavli Institute for the Physics and Mathematics of the Universe (Kavli IPMU, WPI), The University of Tokyo, Chiba 277-8582, Japan}

\date{\today}

\begin{abstract}
We examine the capability of pulsar timing arrays (PTAs) to detect
very small-scale clumps of dark matter (DM), which are a
natural outcome of the standard cold dark matter (CDM) paradigm. A
clump streaming near the Earth or a pulsar induces an impulsive acceleration
to encode residuals on pulsar timing data.   
We show that, assuming the standard abundance of DM clumps predicted
by the CDM model, small-scale DM clumps with masses from 
$\sim 10^{-11} M_\odot$ to $\sim 10^{-8} \ M_\odot$ can be detectable
by a PTA observation for a few decades with ${\cal O}(100)$ of pulsars
with a timing noise of ${\cal O}(10)$~ns located at $\gtrsim 3$~kpc
away from the Galactic center, as long as these mass scales are larger
than the cutoff scale of the halo mass function that is determined by
the particle nature of DM. Our result suggests that PTAs can provide a 
unique opportunity for testing one of the most fundamental predictions
of the CDM paradigm. In addition, the detections and non-detections
can constrain the cutoff mass scale inherent to the DM model.
\end{abstract}

\maketitle


\section{\label{sec:intro}Introduction}

\begin{figure}
\subfigure{\includegraphics[width=75mm]{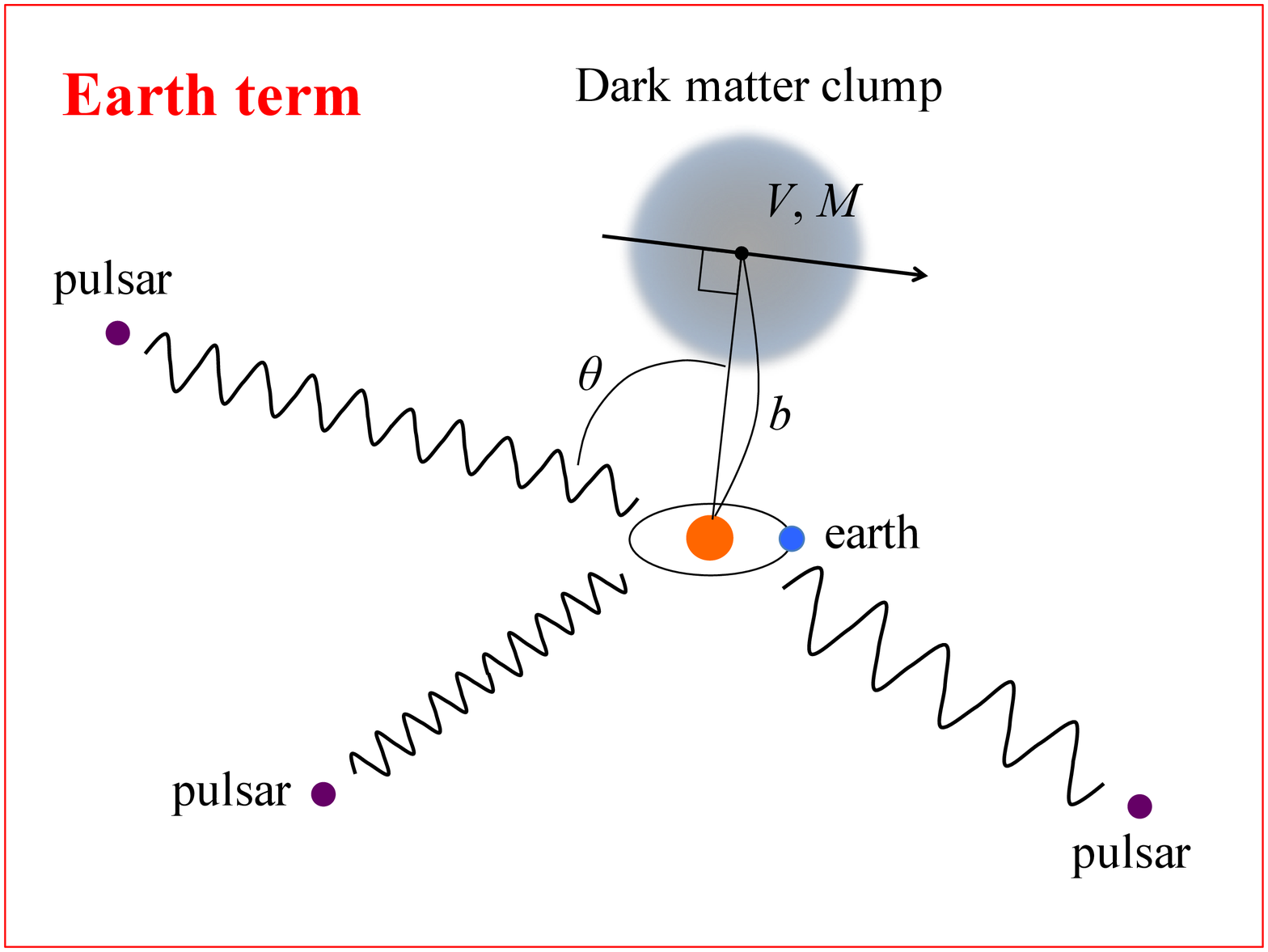}}
\subfigure{\includegraphics[width=75mm]{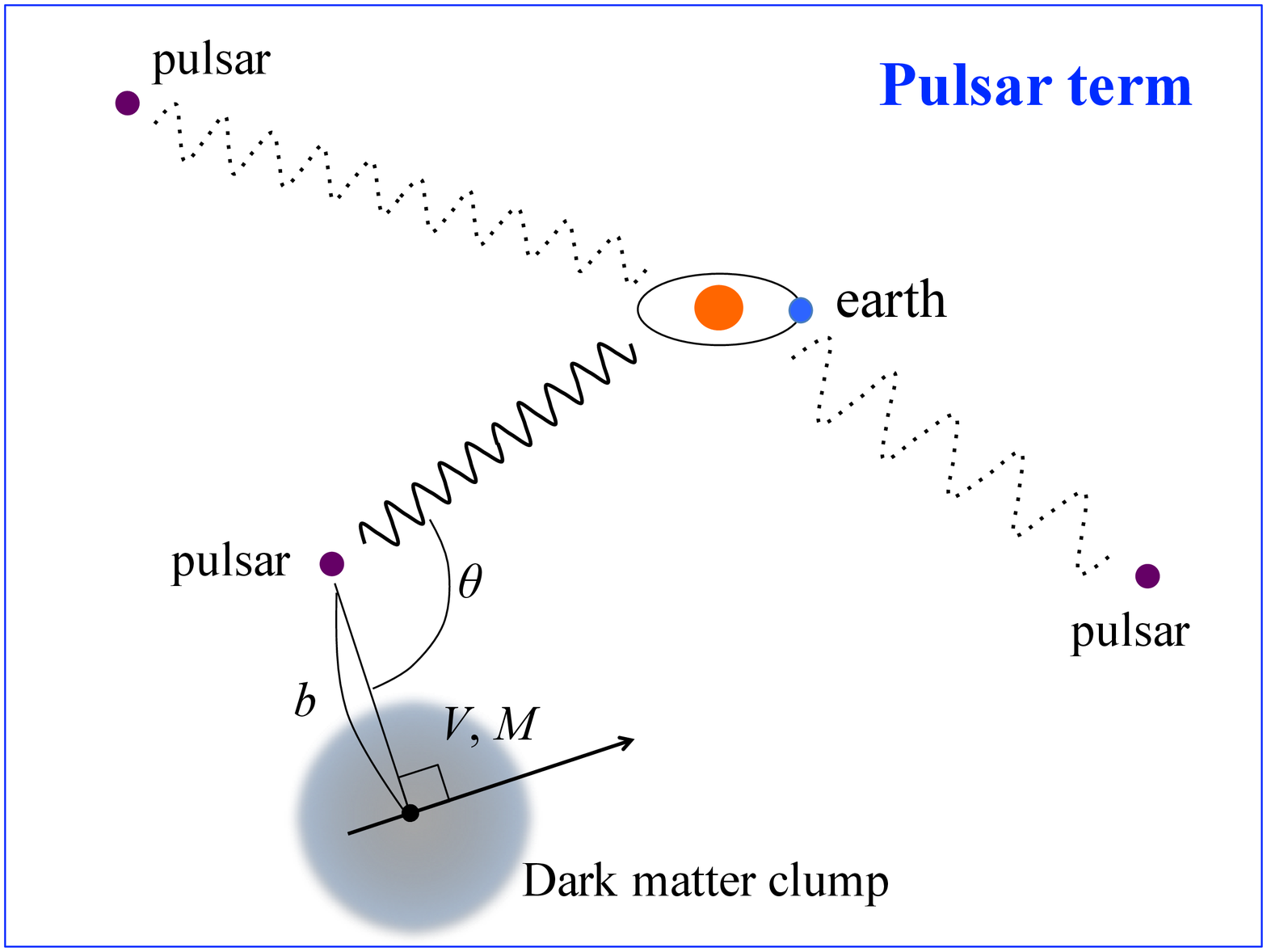}}
\caption{
Schematic pictures of detecting small-scale dark matter (DM) clumps
using a pulsar timing array (PTA).  The impulsive acceleration of a DM
clump is determined by its mass $M$, the DM clump velocity $V$
relative to the target, the impact parameter $b$, and the angle
$\theta$ between the line connecting a pulsar and the Earth and the
line connecting the Earth or pulsar and the DM clump at the closest
approach.  The case with a DM clump passing near the Earth is shown in
the upper panel, whereas the case with a DM clump passing near a
pulsar is shown in the lower panel.
}\label{fig:schem}
\end{figure}

Variety of cosmological observations, e.g., cosmic microwave
background and large-scale structure, can be consistently explained by
the standard $\Lambda$ cold dark matter ($\Lambda$CDM) model with an
initial condition set by inflation.  One of the most fundamental
predictions of the CDM paradigm is that cosmic structure grows
hierarchically such that smaller structures form earlier.
As a consequence, the CDM model predicts the mass function of DM clumps
(halos) that extends to very small masses.  
The cutoff of the mass function at the small mass end is closely
related to the nature of DM, e.g., it corresponds to the free
streaming or kinetic decoupling scale and hence is related to e.g., the
mass of DM particles for weakly interacting massive particle (WIMP)
scenarios 
(e.g., \cite{Jungman:1995df,Berezinsky:2014wya,Bringmann:2009vf}). 
Therefore detections of such small-scale DM clumps are crucial for the
confirmation of the CDM paradigm, yet it is very challenging
observationally.  


Are there any ways to detect such small-scale DM clumps
observationally? One possible way is to detect them indirectly via
annihilations of DM particles (e.g.,
\cite{Berezinsky:2003vn,Diemand:2005vz,Springel:2008zz,Ando:2009fp,Bovy:2009zs,Ishiyama:2010es}).
However, no DM annihilation signal was detected by the Fermi
Gamma-ray Space Telescope, from which tight constraints on the DM
annihilation cross section has been obtained \cite{Ackermann:2015zua}. 
Another possible way is microlensing, although it appears that it
would be difficult to probe very small ($M\lesssim 100~M_\odot$) mass
clumps by microlensing (e.g., \cite{Erickcek:2010fc}).


In this paper, we propose to utilize pulsar timing arrays (PTAs) for
searching small-scale DM clumps in the Galaxy. Previous studies
considered the Shapiro time delay that is induced when an ultracompact
DM halo crosses the line-of-sight between the Earth and a pulsar 
\cite{Siegel:2007fz,Baghram:2011is,Clark:2015sha,Clark:2015tha}. 
Also PTA has been shown to be able to detect oscillations of
gravitational potential induced by ultralight scalar DM with mass
around $10^{-23}-10^{-22} \ \rm eV$~\cite{Khmelnitsky:2013lxt}.  
Instead, we consider an impulsive acceleration of the Earth or one of the
pulsars that is induced by a DM clump streaming near the Earth or the
pulsar (Fig.~\ref{fig:schem}), which was originally proposed to detect
point mass objects such as primordial black holes
(PBHs)~\cite{Seto_Cooray_07,KK_Seto_12}.
The idea of using PTAs to probe small-scale DM clumps has been
briefly mentioned in \cite{Ishiyama:2010es}. In this paper, we show
that PTAs can provide independent and interesting constraints on the
abundance of small-scale DM clumps with masses from $\sim 10^{-11}
\ M_\odot$ to $\sim 10^{-8} M_\odot$.   

This paper is organized as follows. In Sec.~\ref{sec:clump}, we
estimate the mass distribution of small-scale DM clumps in the
Galaxy. In Sec.~\ref{sec:PTA}, we show the capability of PTAs for
detecting them. Sec.~\ref{sec:discussion} is devoted to discussions.
Throughout the paper, we use cosmological parameter values obtained by
{\it Planck}~\citep{Ade:2015xua}.

\section{Small-Scale Clumps in the Galaxy}\label{sec:clump}
We first consider the formation of small-scale DM clumps in 
Sec.~\ref{sec:formation} and then the destruction processes and
survival probability in Sec.~\ref{sec:destruction}.

\subsection{Formation}\label{sec:formation}
In the concordance cosmology, seeds of structures are fluctuations
generated during inflation. The primordial power spectrum is nearly
scale-invariant with a small amplitude of ${\cal O} (10^{-5})$.
The matter fluctuations start to grow linearly with the scale factor
at around the matter-radiation equality, $z \sim z_{\rm eq} = 2.35
\times 10^4 \ \Omega_{\rm m} h^2$.  When the amplitude of a
fluctuation becomes ${\cal O}(1)$, it collapses into a halo or clump
dominated by DM. In the spherical collapse model, this happens at 
\begin{equation}
1+ z \approx \left[ \frac{\nu \sigma_{\rm eq} (M)}{\delta_{\rm c}}\right] (1 + z_{\rm eq}), 
\end{equation}
for a clump with mass $M$ formed from a $\nu$-sigma fluctuation.
Here $\sigma_{\rm eq} (M)$ is the variance of the linear density
fluctuation on a scale corresponding to a mass $M$ at the
matter-radiation equality, and $\delta_{\rm c}=(3/20)(12\pi)^{2/3}$. 
The variance $\sigma_{\rm eq}(M)$ at mass scales larger than 
$M\sim 10^{12}M_\odot$ has been well determined observationally 
(e.g., \cite{Hlozek:2011pc}). We can extrapolate the variance to
smaller mass scales by assuming a power-law primordial power
spectrum ($\propto k^{n_{\rm p}}$), which is a reasonable assumption
under inflationary scenarios. In this case, we may use an analytical
fitting formula given in \cite{Berezinsky:2003vn} (see their Eqs.~92-94)
\begin{equation}\label{eq:sigma_eq}
\sigma_{\rm eq}(M) \approx \frac{2\times 10^{-4}}{\sqrt{f_s(\Omega_\Lambda)}}\left[-\frac{1}{3}\ln\left(\frac{M}{M_{h, \rm eq}}\right)\right]^{3/2}\left(\frac{M}{M_{h, 0}}\right)^{-(n_p-1)/6},
\end{equation}
where $M_{h, \rm eq} \approx 7.5 \times 10^{15}  \ \Omega_{\rm m}{}^{-2} h{}^{-4} \ M_\odot$ and $M_{h,0} \approx 3.0 \times 10^{22} \ \Omega_{\rm m} h^{-1}  \ M_\odot$
are the mass inside the cosmological horizon at $z = z_{\rm eq}$ and $z = 0$, respectively, and
$f_s(\Omega_\Lambda) = 1.04 - 0.82 \Omega_\Lambda + 2 \Omega_\Lambda{}^2$.
The variance $\sigma_{\rm eq} (M)$ is a decreasing function of mass
$M$, indicating that structures with smaller masses form earlier.


We then estimate the density and radius of DM clumps at their formation.
In the spherical collapse model (e.g., \cite{Lacey:1993iv}), the
average density can be given as  
\begin{equation}\label{eq:barrho}
\bar \rho \approx \kappa \rho_{\rm eq} 
\left[ \frac{\nu \sigma_{\rm eq}(M)}{\delta_{\rm c}}\right]^3,
\end{equation}
where $\kappa = 18 \pi^2$, 
$\rho_{\rm eq} = \rho_{0}(1+z_{\rm eq})^3$,
and $\rho_0 = 2.78 \times 10^{-7} \ \Omega_{\rm m} h^2 \ M_\odot {\rm 
  \ pc^{-3}}$ is the mean matter density of the universe at $z = 0$.
The characteristics radius is 
\begin{equation}\label{eq:barR}
\bar R \approx \left(\frac{3M}{4\pi \bar \rho}\right)^{1/3}.
\end{equation}
Using Eq.~(\ref{eq:sigma_eq}), Eqs.~(\ref{eq:barrho}) and
(\ref{eq:barR}) for a range of masses $10^{-12} M_\odot \lesssim M
\lesssim 10^{-6} \ M_\odot$ 
the average density can be approximated as
\begin{equation}\label{eq:barrho_sub}
\bar \rho \sim 0.41 \  M_\odot \ {\rm pc^{-3}} \ \left(\frac{\nu}{2}\right)^3 \left(\frac{M}{10^{-10} \ M_\odot} \right)^{-3\alpha}, 
\end{equation}
and the radius of the clump is 
\begin{equation}\label{eq:barR_sub}
\bar R \sim 83 \ {\rm AU} \ \left(\frac{\nu}{2}\right)^{-1} \left(\frac{M}{10^{-10} \ M_\odot}\right)^{\alpha+1/3}, 
\end{equation}
where $\alpha \approx 0.02$. 

We assume that primordial fluctuations are nearly
scale-invariant and follow the Gaussian distribution, which are
consistent with observations at large scales. In this case, the clump
formation probability is $\propto e^{-\nu^2/2}$, indicating that
clumps from low-sigma fluctuations dominates. 
Also, the average density of the clumps are almost scale invariant;
larger mass clumps are only slightly denser (see
Eq.~\ref{eq:barrho_sub}).  However, a non-negligible amount of
high-sigma fluctuations ($\nu \gg 1$) may be seeded for a certain
range of clump mass, depending on inflation models.  
Such clumps, the so-called ultracompact minihalos (UCMHs), can have
larger survival probabilities and enhanced observational signatures 
(e.g., \cite{Berezinsky:2003vn,Ricotti:2009bs,Scott:2009tu}).  
In this paper, we do not consider such possibility and focus on
``normal'' DM clumps from low-sigma fluctuations, which should
represent a conservative estimate of the detectability of small-scale
DM clumps. 

Regardless of inflation models, there should be a lower cut-off mass
scale of the clump formation. For instance, thermal relic DM models
have free-streaming scales, below which clustering of DM particles is
suppressed by their streaming motions. Damping of the power spectrum
is also induced by acoustic oscillations after kinetic decoupling.
In the case of 100~GeV WIMP model, the cutoff is located at around
$M_{\rm cutoff}\sim 10^{-6} \ M_\odot$ (e.g., \cite{Green:2005fa}),
although we note that the 100~GeV thermal relic WIMP model appears to
be already excluded by Fermi $\gamma$-ray observations \cite{Ackermann:2015zua}. 
More massive ($>$TeV) WIMP models predict small-scale DM clumps as small as $\sim 10^{-10}M_\odot$
\cite{Bringmann:2009vf}. Furthermore, non-thermal DM models such as
axions can have critical scales qualitatively similar to the
above. For example, in the case of 10~$\mu$eV axion, the cutoff mass
scale is at $M_{\rm cutoff} \sim 10^{-12} \ M_\odot$ (e.g., \cite{Marsh:2015xka}). 
One can strongly constraint the nature of DM by detecting these cutoff scales.

\subsection{Destruction and survival probability}\label{sec:destruction}
Next we estimate the current abundance of small-scale DM clumps in the
Galaxy. Since these DM clumps are fluffy as seen from
Eqs.~(\ref{eq:barrho_sub}) and (\ref{eq:barR_sub}), they can be
destructed due to gravitational interactions within the Galaxy.   
The most relevant process is the tidal destruction~\cite{1972ApJ...176L..51O}:  
a clump will be destructed when the sum of the internal energy induced
by tidal shocks,  
\begin{equation}
\delta E = \frac{3-\beta}{6(5-2\beta)} M \sum_i \delta v_i{}^2
\label{eq:shock}
\end{equation}
becomes comparable to the binding energy of the clump
\begin{equation}
E_{\rm b} = \frac{3-\beta}{2(5-2\beta)}\frac{GM^2}{R},
\label{eq:bind}
\end{equation}
where
\begin{equation}
\delta v_i = \frac{2b_ig_i}{V_i}
\end{equation}
is the tidally-induced velocity perturbation in a clump experiencing
an encounter with a gravitational source with a impact parameter
$b_i$, the relative velocity $V_i$, and the tidal force per unit mass
$g_i$, and $\beta$ represents the power-law index of the density
profile of the clump at the characteristic radius, i.e.,  
\begin{equation}
\rho = \frac{3 - \beta}{3} \bar \rho \left(\frac{R}{\bar R}\right)^{-\beta}.  
\end{equation}
In this paper, we set $\beta = 2$ with noting that our results are not
sensitive to $\beta$ as long as $1 \lesssim \beta \lesssim 2$.  
As in the previous
studies~\cite{Berezinsky:2003vn,Berezinsky:2005py,Berezinsky:2006qm,Berezinsky:2007qu},  
we divide the destruction history into two parts; during the
hierarchical structure formation of the Milky-Way halo and after the
star formation in the Galaxy. 

In the course of forming hierarchical structures, DM clumps can merge
and be destructed by the tidal interaction between with each other.  
The DM clump mass density relative to the total DM mass density in the
end is calculated as \cite{Berezinsky:2003vn} 
\begin{equation}\label{eq:xi_M_nu}
\xi (M, \nu) d\nu \frac{dM}{M} \approx y(\nu) \frac{e^{-\nu^2/2}}{\sqrt{2\pi}} (n+3) d\nu \frac{dM}{M}, 
\end{equation}
where $y(\nu)$ is a monotonically increasing function of $\nu$ (see
Fig. 2 of \cite{Berezinsky:2003vn}), representing the fact that less
dense clumps formed from a smaller fluctuation are selectively
disrupted, and  
\begin{equation}\label{eq:n_clump}
n = -3\left[1+2\frac{\partial\log\sigma_{\rm eq}(M)}{\partial\log M}\right].
\end{equation}
Integrating Eq.~(\ref{eq:xi_M_nu}) over $\nu$, the fraction of DM mass in clumps with a mass $M$ can be given as  
\begin{equation}\label{eq:xi_i}
\xi_{\rm i} \frac{dM}{M} \simeq 0.02 (n+3)\frac{dM}{M}.
\end{equation}
It is worth noting that surviving clumps are mainly from low-sigma
fluctuations, i.e., $\nu \lesssim 2$ \cite{Berezinsky:2003vn}.  

The calculated $\xi_i$ from Eqs.~(\ref{eq:sigma_eq}),
(\ref{eq:n_clump}), and (\ref{eq:xi_i}) is shown in Fig.~\ref{fig:xi}
(dotted line).  The clump mass density depends little on the clump
mass even after taking account of the destruction process, 
$\xi_i \propto M^0$. This is due to the fact that the survival
probability is sensitive to the average clump density, which is also
almost scale invariant, $\bar \rho \propto M^0$ (see Eq.~\ref{eq:barrho_sub}).  
At this point, $\gtrsim 0.1 \%$ of the DM mass are in the form of
small-scale clumps in each logarithmic mass interval $d \ln M$.
Numerical simulations of the hierarchical structure formation predict
clump abundances that are consistent with the estimate described above
(e.g., \cite{Diemand:2005vz,Springel:2008cc,Ishiyama:2014uoa,Stucker:2017nmi}), 
although the uncertainties are still large.
In Fig.~\ref{fig:xi}, we also show the impact of the cutoff in the
mass function, which is assumed to be an exponential cutoff ($\propto
\exp[-(M_{\rm cutoff}/M)^{2/3}]$) for simplicity.

When the Galaxy is formed at the center of the halo, DM clumps are
further destructed by stars in the Galaxy. For example, in the case of
a clump passing near a star with mass $M_\ast$, the exerted tidal
force is on average $g \approx GM_\ast \bar R/b^3$,  
and it is tidally destructed if $b \lesssim b_{\rm c}$ where 
\begin{eqnarray}\label{eq:b_c}
b_{\rm c} \sim 3000 \  &{\rm AU}& \ \left( \frac{\nu}{2}\right)^{-3/4} \left(\frac{M}{10^{-10} \ M_\odot}\right)^{3\alpha/4} \notag \\
&\times& \left(\frac{V}{350 \ {\rm km \ s^{-1}}}\right)^{-1/2} \left(\frac{M_\ast}{M_\odot}\right)^{1/2}.
\end{eqnarray}
We should note that the relative velocity is typically much larger
than the tidally induced velocity, i.e., $\delta v/V \ll 1$, even for
$b \lesssim b_{\rm c}$.  This means that the bulk of the destructed
clump will not be captured by the star but fly away as a DM stream.

Here we estimate the surviving probability of clumps against the tidal
destruction by Galactic stars following \cite{Berezinsky:2006qm}. 
To this end, one has to assume the surface density of the stellar disk
($\sigma_{\rm d}$) and the DM halo profile of the Milky Way
($\rho_{\rm h}$).  The former determines the survival probability of a
clump per disk crossing,  while the latter sets the orbits of clumps. 
Here we assume 
\begin{equation}
\sigma_{\rm d}(r) = \frac{m_{\rm d}}{2 \pi r_{\rm d}{}^2} e^{-r/r_{\rm d}}
\end{equation}
with $m_{\rm d} = 6 \times 10^{10} \ M_\odot$ and $r_{\rm d} = 2.6 \ \rm kpc$~\cite{Nesti:2013uwa}
and the NFW profile
\begin{equation}
\rho_{\rm h}(r) = \frac{\bar \rho_{\rm h}}{(r/L)(1+r/L)^2}
\end{equation}
with $\bar \rho_{\rm h} = 1.4 \times 10^{7} \ M_\odot \ \rm kpc^{-3}$
and $L = 16 \ \rm kpc$.   
The resultant DM mass fraction in clumps in the Galaxy are shown in
Fig.~\ref{fig:xi}.  The thick and thin solid black lines indicate the
cases at the Galactic radii of $8 \ \rm kpc$ and $3 \ \rm kpc$,
respectively. Compared with $\xi_i$ (Eq.~\ref{eq:xi_i}), the survival
fraction of DM clumps is decreased by $\gtrsim 10 \ \%$ at around the
solar system.  On the other hand, more than $99 \ \%$ of the DM clumps
have been destructed at an inner radius where stars are more densely
distributed.  Again, the survival probability does not depend strongly
on the clump mass. Our results are broadly consistent with previous
studies~\cite{Berezinsky:2003vn,Berezinsky:2005py,Berezinsky:2006qm,Berezinsky:2007qu}. 

\section{Searching dark matter clumps with pulsar timing arrays}\label{sec:PTA}

\begin{figure}
\includegraphics[width=90mm]{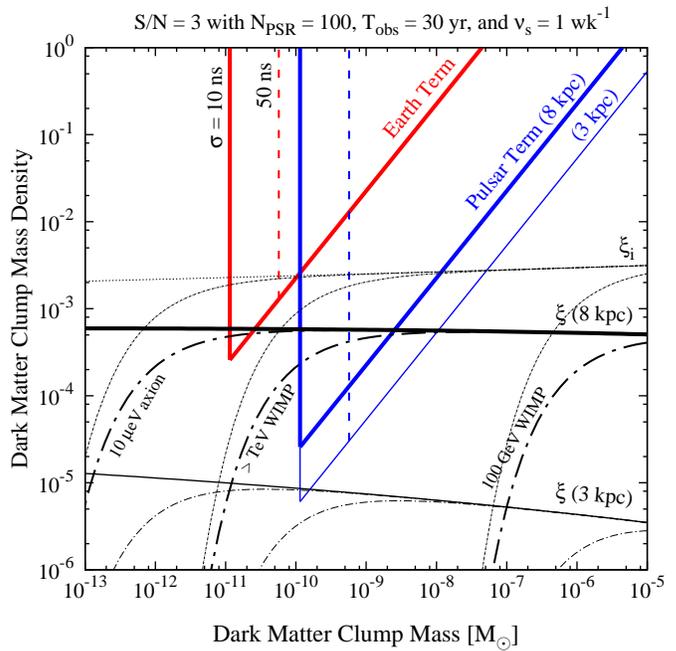}
\caption{
The dark matter (DM) clump mass density relative to the total DM mass
density in the Galaxy and the sensitivity of a pulsar timing array.   
The thick and thin black solid lines show the current mass density
distribution of DM clumps at 8 and 3~kpc from the Galactic center,
respectively.  The dotted line indicates the mass density distribution
without the effect of Galactic stars (see Eq.~\ref{eq:xi_i}).  
The cutoffs depending on DM models are shown with dotted-dash
lines. The red and blue lines indicate the parameter region in which
DM clumps are detectable using the Earth and pulsar terms,
respectively. For the pulsar term, the thick and thin lines show the
cases that all the pulsars locate at 8 and 3~kpc from the Galactic
center, respectively. We assume a future PTA observation with $\sigma
= 10 \ \rm ns$ ({\it solid}), $50 \ \rm ns$ ({\it dashed}),  
and $\nu_{\rm s} = 1 \ {\rm wk^{-1}}$, $N_{\rm PSR} = 100$, $T_{\rm obs} = 30
\ \rm yr$, and set the detection threshold as ${\rm S/N}=3$.   
}\label{fig:xi}
\end{figure}

We now discuss how to detect small-scale DM clumps in the Galaxy with
PTAs, following \cite{KK_Seto_12} in which the possibility of
constraining the abundance of PBHs with PTAs was explored.  The method
is schematically described in Fig.~\ref{fig:schem}. 
We consider an impulsive acceleration driven by a flyby of a DM clump
around a target mass, either the Earth or one of the pulsars. 
Residuals induced in the timing-of-arrival (TOA) data by the
former and latter events are called Earth and pulsar terms,
respectively.  

Hereafter, we consider a PTA that consists of $N_{\rm PSR}$ pulsars with a comparable timing noise $\sigma$. 
For simplicity, we assume the noises are white and have no correlation. 
The observation is also characterized by the total duration $T_{\rm obs}$ 
and sampling rate of the timing-of-arrival (TOA) data
$\nu_{\rm s}$. For example, the Parkes PTA currently consists of
$N_{\rm PSR} \sim 20$ of pulsars with timing precisions of $\sigma
\sim 100 \ \rm ns$,  and the conventional data sampling rate is
$\nu_{\rm s} \sim 0.5 \ \rm wk^{-1}$ \cite{Reardon:2015kba}. 
These parameters could be significantly improved 
(i.e., larger $N_{\rm PSR}$, smaller $\sigma$, and larger $\nu_{\rm s}$) in
the era of Square Kilometre Array (but see e.g.,
\cite{Shannon:2014kha,Cordes:2015hia}).  

We first describe the signal of a flyby event. 
The impulsive acceleration of the Earth or a pulsar occurs with a timescale of $T = b/V$, or 
\begin{equation}\label{eq:sig_dur}
T \sim 30 \ \mathrm{yr} \left( \frac{b}{2200 \ \mathrm{AU}}\right)  \left( \frac{V}{350 \ \mathrm{km/s}}\right)^{-1},
\end{equation}
inducing timing residuals mainly at a Fourier frequency $f = 1/T$. 
The mode amplitude is given as $s_f \approx (a T^2/c) \times |
\cos\theta |$, where $a\approx GM/b^2$ is the acceleration 
and $\theta$ is the angle defined in Fig.~\ref{fig:schem}. 
Then, we can estimate the amplitude as 
\begin{equation}\label{eq:sig_amp}
s_f \sim 0.20 \ \mathrm{ns} \left( \frac{M}{10^{-10} \ M_\odot}\right) \left( \frac{V}{350 \ \mathrm{km/s}}\right)^{-2} \left( \frac{| \cos \theta |}{0.58} \right).
\end{equation}
In the above estimate, we assume that the clump is a point mass object.
As we will see below, the maximum detectable impact parameter $b_{\rm
  max}$ is typically comparable to $b_{\rm c}$ in Eq.~(\ref{eq:b_c}),  
which means that detected clumps are being destructed by the Sun or
one of the pulsars, leading the modification of the signal from the
point mass approximation.  However, this finite size effect is minor
given that the tidally induced velocity is much smaller than the flyby
velocity, i.e., $\delta v/V \ll 1$.   

Neglecting the finite size effect, the signal of a DM clump is
characterized by $T$ and $s_f$.  On the other hand, we have four
physical parameters, $M$, $b$, $V$, and $\theta$,  
which reflect the abundance and dynamics of DM clumps in the Galaxy. 
In principle, the degeneracies in the parameters can be broken
statistically. Also, the orbit of a DM clump approaching to the Earth
can be estimated from the dipole pattern of the Earth term.  
Hereafter, we set fiducial values of $V$ and $\theta$ as follows. 
Assuming that scattering between DM clumps and the targets is
isotropic, the ensemble average of $\theta$ reduces to $\sqrt{ \langle
  \cos^2\theta \rangle } = 1/\sqrt{3} \sim 0.58$. We assume that
typical value of $V$ is equal to the rms velocity of DM relative to
the solar system, $\sim 350 \ \rm km \ s^{-1}$, which was estimated
dynamically \citep{Carr_Sakellariadou_99}.  
Give that the typical peculiar velocity of the observed millisecond
pulsars are relatively small $\lesssim 100 \ \rm km \ s^{-1}$ with
some exceptions \citep{Hobbs:2005yx}, it is reasonable to adopt $V =
350 \ \rm km \ s^{-1}$ as a fiducial value for both the Earth and
pulsar terms.   

Next, we evaluate the timing noise of the detector. The Fourier mode
of the noise at the frequency $f=1/T$ can be estimated as $\approx
\sigma/\sqrt{T \nu_{\rm s}}$ for a pulsar term. When using the Earth
terms, the noise can be statistically reduced by a factor of $\approx
1/\sqrt{N_\mathrm{PSR}}$. As a result, the effective noise for
the DM clump search can be described as
\begin{eqnarray}\label{eq:noise}
n_f \sim 0.25 \ &\mathrm{ns}& \left( \frac{\sigma}{10 \ \mathrm{ns}} \right) \left( \frac{T}{30 \ \mathrm{yr}} \right)^{-1/2} \left( \frac{\nu_{\rm s}}{1 \ \mathrm{wk}^{-1}} \right)^{-1/2} \notag \\ 
&\times& N_\mathrm{PSR}{}^{-E/2}
\end{eqnarray}
with $E = 0 \ \mathrm{or} \ 1$ for the pulsar and Earth terms,
respectively.  From Eqs.~(\ref{eq:sig_amp}) and (\ref{eq:noise}), the
signal-to-noise ratio  is given as ${\rm S/N} \equiv s_f/n_f$.  

Finally, the event rate of DM clumps with a mass $M$ passing near the
Earth with an impact parameter $b$ is $\approx \pi b^2 V
\xi(M,r)\rho_{\rm h}(r)/M$.  Such flybys occur in total $N_{\rm PSR}$
times more for the pulsars.  
Accordingly, the event rate can be estimated as 
\begin{eqnarray}\label{eq:rate}
{\cal R} \sim 0.014 \ &{\rm yr^{-1}}& \left(\frac{\xi}{10^{-3}}\right) \left(\frac{\rho_{\rm h}}{0.011 \ M_\odot \ \rm pc^{-3}}\right) \notag \\ 
&\times& \left(\frac{M}{10^{-10} \ M_\odot}\right)^{-1}  \left(\frac{b}{2200 \ \rm AU}\right)^{2} \notag \\
&\times&  \left(\frac{V}{350 \ \rm km/s}\right)  N_{\rm PSR}{}^{1-E}. 
\end{eqnarray}

Detections of small-scale DM clumps by a PTA observation is possible if 
(i) the signal-to-noise ratios are sufficiently large, e.g., 
${\rm S/N}> 3$, (ii) the signal duration is shorter than the
duration of the observation, $T < T_{\rm obs}$, and  (iii) more than
one events occur within the duration of the observation, ${\cal R}T_{\rm obs} > 1$.  
From Eqs.~(\ref{eq:sig_dur})-(\ref{eq:rate}), these conditions can be
translated into  
\begin{eqnarray}\label{eq:det_con_i}
\left(\frac{M}{10^{-10} \ M_\odot}\right)\left(\frac{b}{2200 \ {\rm AU}}\right)^{1/2} &\gtrsim& 1.1 \left(\frac{\sigma}{10 \ {\rm ns}}\right)\left(\frac{\nu_{\rm s}}{1 \ {\rm wk^{-1}}}\right)^{-1/2} \notag \\
&\times& \left(\frac{V}{350 \ {\rm km/s}}\right)^{5/2} \notag \\
&\times& \left(\frac{{\rm S/N}}{3}\right) N_{\rm PSR}{}^{-E/2}, 
\end{eqnarray}
\begin{eqnarray}\label{eq:det_con_ii}
b \lesssim b_{\rm max} = 2200 \ {\rm AU} \left(\frac{T_{\rm obs}}{30 \ {\rm yr}}\right)\left(\frac{V}{350 \ {\rm km/s}}\right), 
\end{eqnarray}
and
\begin{eqnarray}\label{eq:det_con_iii}
\left(\frac{M}{10^{-10} \ M_\odot}\right)\left(\frac{b}{2200 \ {\rm AU}}\right)^{-2} &\lesssim& 0.43 \left(\frac{\xi}{10^{-3}}\right)\notag \\
&\times& \left(\frac{\rho_{\rm h}}{0.011 \ M_\odot \ {\rm pc^{-3}}}\right)  \notag \\
&\times& \left(\frac{V}{350 \ {\rm km/s}}\right)\left(\frac{T_{\rm obs}}{30 \ {\rm yr}}\right)  \notag \\
&\times& N_{\rm PSR}{}^{1-E},
\end{eqnarray}
respectively. We note that the conditions (i) and (iii)
(Eqs.~\ref{eq:det_con_i} and \ref{eq:det_con_iii}) are more easily
satisfied for larger $b$, because both S/N and ${\cal R}$ become
larger. As a result, the minimum detectability is always set by cases
with $b \sim b_{\rm max}$ in Eq.~(\ref{eq:det_con_ii}), which is at
most a few 1000~AU for an observation within a human timescale. The
detectable parameter range of DM clumps with a given PTA observation
can be obtained by substituting $b = b_{\rm max}$ into
Eqs.~(\ref{eq:det_con_i}) and (\ref{eq:det_con_iii}).  

In Fig.~\ref{fig:xi}, the red and blue lines show the detectable
parameter regions for a PTA observation with $\nu_{\rm s} = 1 \ {\rm
  wk^{-1}}$, $N_{\rm PSR} = 100$, and $T_{\rm obs} = 30 \ \rm yr$,
setting the detection threshold to ${\rm S/N}= 3$. 
For the solid and dashed lines, the average timing noises are assumed
to be $\sigma = 10$ and $50 \ \rm ns$, respectively. The minimum clump
mass is set by Eqs.~(\ref{eq:det_con_i}) and (\ref{eq:det_con_ii})
while the minimum mass density is set by Eqs.~(\ref{eq:det_con_ii})
and (\ref{eq:det_con_iii}). 

We find that the Earth and pulsar terms are highly complementary \cite{KK_Seto_12}. 
One can detect $\sim 1/\sqrt{N_\mathrm{PSR}}$ times lighter clumps with
the Earth term and $\sim N_\mathrm{PSR}$ times less abundant clumps
with the pulsar term, which is optimal for more massive clumps.  
Note that the effective sensitivity of using the pulsar term depends
on the local DM density (see Eq.~\ref{eq:det_con_iii}) and thus the
location of the pulsars in the Galaxy.  The thick and thin blue lines
correspond to the cases where all the pulsars are located at 8~kpc and
3~kpc from the Galactic center, respectively.  

From Fig.~\ref{fig:xi}, we conclude that small-scale DM clumps with
masses ranging from $\sim 10^{-11} \ M_\odot$ to $\sim 10^{-8}
\ M_\odot$ predicted by the standard CDM model can be detected by a
PTA observation for a few decades with ${\cal O}(100)$ of pulsars with
a timing noise of ${\cal O}(10)$~ns located at $\gtrsim 3$~kpc away
from the Galactic center, if the cutoff mass scale is lower than these
masses. Therefore the detection and non-detection of small-scale DM
clumps by such PTA observations can provide interesting constraints on the
particle nature of DM.

\section{\label{sec:discussion}Discussion}
Our work has demonstrated that small-scale DM clumps predicted in the
standard framework of the CDM model can be detected by a PTA
observation. Our estimate is based on a reasonable extrapolation of
the well-constrained CDM model to small scales, and is based on
conservative assumptions. For instance, recent work by
\cite{vandenBosch:2017ynq,vandenBosch:2018tyt} has shown that our
condition of the destruction of DM clumps by the 
ratio tidal shock (Eq.~\ref{eq:shock}) to the binding energy
(Eq.~\ref{eq:bind}) grossly overestimates the fraction of clumps that
undergo the destruction. This suggests that the abundance of DM
clumps derived in this paper, and therefore the event rate of the
impulsive acceleration detectable by PTAs, may be significantly
underestimated. 

However, a caveat is its observational challenge. Given the unresolved
systematic noises in timing data (e.g.,
\cite{Shannon:2014kha,Cordes:2015hia}), the proposed PTA observation
to detect the DM clump density predicted by the standard CDM model
will be challenging even in the era of Square Kilometre
Array. Therefore development of new methodology to further reduce
systematic noises in timing data is highly anticipated. 

While we have considered the detectability of small-scale DM clumps in
a standard CDM scenario, depending on inflationary models the
primordial power spectrum can be more enhanced towards small scales,
leading to more abundant DM clumps or UCMHs. For such cases, a more
conservative PTA configuration still provides meaningful constraints.  
Although some search methods have been already proposed for UCMHs 
(e.g.,
\cite{Siegel:2007fz,Baghram:2011is,Clark:2015sha,Clark:2015tha}), our
approach is complementary to these previous approaches. 
In particular, we note that another PTA method using the Shapiro time
delay is sensitive to larger DM clump masses of $M \gtrsim 10^{-4}
\ M_\odot$~\cite{Siegel:2007fz,Clark:2015sha}. 

We should also discuss competitive sources for a robust detection of
DM clumps with PTAs. First of all, gravitational waves (GWs) from
binary super-massive black holes are the primary target of PTAs~(e.g.,
\cite{Sesana:2008xk,Sesana:2008mz}).  The amplitude and duration of
the signal will be comparable to those of DM clumps. In principle, the
signals can be distinguished by the multipole patterns in the Earth
term as GWs are quadrupole whereas flyby events are dipole.  

A possible concern is that flyby events of non-DM objects, e.g.,
floating planets, may serve as false positives of DM clump signals. 
However, the expected event rate for these non-DM objects is
significantly smaller than that of DM clumps.  Also, in the case of
using the Earth term, non-DM objects may be identified by other
instruments such as optical telescopes.  
 
Our method was originally proposed to detect PBHs with $\sim 10^{-11}
- 10^{-8} \ M_\odot$~\cite{Seto_Cooray_07,KK_Seto_12}.  The search for
microlensing with the Subaru Hyper Suprime-Cam provides independent
tight constraints on the abundance of PBHs in a similar mass
range~\cite{Niikura:2017zjd}. Since small-scale DM clumps are much
less sensitive to microlensing due to their fluffy nature, the
combination of microlensing searches and PTAs will allow us to
statistically disentangle signals by PBHs and small-scale DM clumps in
PTA observations. Furthermore, for high S/N events, flybys of DM clumps
and PBHs might be distinguished by the details of the signals, because
only the formers are destructed by the tidal shock.  

Finally, if an ultralight scalar field with mass around $10^{-23}
-10^{22} \ \rm eV$ constitutes a good fraction of the DM, stochastic
oscillations of gravitational potential can be detectable by future
PTAs \cite{Khmelnitsky:2013lxt}.  Although the amplitude and duration
of the signal can be also comparable to those of DM clumps, it is
expected to be monochromatic and induce monopole patterns in the Earth
term. Thus such signals can be distinguished from flyby events of DM
clumps.

\section*{Acknowledgement}
The authors thank George Hobbs, Teruaki Suyama and Shin'ichiro Ando
for fruitful discussion. This work is partially supported by JSPS
KAKENHI Grant Number JP17K14248, JP26800093, JP15H05892, and WPI
Initiative, MEXT, Japan. 

\bibliographystyle{apsrev4-1}
\bibliography{ref}

\end{document}